\documentclass[aps,prb,manuscript,showpacs]{revtex4}
\usepackage{amsmath,bm,graphicx}

\def\({\left(}
\def\){\right)}
\def\[{\left[}
\def\]{\right]}

\begin{document}

\title{Effects of  inhomogeneous broadening on reflection spectra of
Bragg multiple quantum well structures with  a defect}

\author{L. I. Deych}
\author{M. V. Erementchouk}
\author{A. A. Lisyansky}

\affiliation{Physics Department, Queens College, City University
of New York, Flushing, New York 11367}

\begin{abstract}
The reflection spectrum of a multiple quantum well structure with
an inserted defect well is considered. The defect  is
characterized by the exciton frequency different from that of the
host's wells. It is shown that for relatively short structures,
the defect produces significant modifications of the reflection
spectrum, which can be useful for optoelectronic applications.
Inhomogeneous broadening is shown to affect the spectrum in a
non-trivial way, which cannot be described by the standard linear
dispersion theory. A method of measuring parameters of both
homogeneous and inhomogeneous broadenings of the defect well from
a single CW reflection spectrum is suggested.
\end{abstract}

\pacs{78.67.De,42.70.Qs,71.36.+c,42.25.Bs} \maketitle

\section{Introduction}

Optical properties of multiple quantum well (MQW) structures have
been attracting a great deal of attention during the past decade.
\cite{Keldysh,IvchenkoSpectrum,CitrinSpectrum,IvchenkoMQW,Andreani1994,Hubner1996,%
Stroucken,Vladimirova,Merle,Haas,Hayes,Khitrova,DLPRBSpectrum,CitrinDefect,DefectMQWOL,DefectMQW}
The main motivation for this interest is the potential in these
systems for  effective control of the light-matter interaction.
Quantum wells in MQW structures are separated from each other by
relatively thick barriers, which prevent a direct interaction
between excitons localized in different wells. They however, can
be coupled by a radiative optical field, and their optical
properties, therefore, become very sensitive to the arrangement of
the structures. Most of the initial research in this area was
focused on properties of periodic MQW's, which consist of
identical quantum wells separated by identical barriers. The
radiative coupling in this case gives rise to MQW polaritons --
coherently coupled oscillations of the exciton polarization and
light. In structures with a small number of wells, the spectrum of
these collective excitations consists of a discrete number of
quasi-stationary (radiative) modes with a re-distributed
oscillator strength: the modes can be classified as super- or
sub-radiant.\cite{CitrinSpectrum,Andreani1994,IvchenkoSpectrum}
With an increase of the number of wells in the structure, the
radiative life-time of the latter decreases, and the life-time of
the former increases. When the number of periods in the structure
becomes large enough, the modes of the MQW are more conveniently
described in terms of stationary polaritons of infinite periodic
structures.\cite{DLPRBSpectrum} The spectrum of polaritons in this
case consists of two branches separated by a polariton gap, which
is normally proportional to the light-exciton coupling constant,
$\Gamma$. There exists, however, a special arrangement of MQW's
where the magnitude of the gap can be significantly increased. If
the period of the MQW structure is made equal to the
half-wavelength of the exciton radiation, the geometric, Bragg,
resonance occurs at the same frequency as the exciton resonance.
As a result, the band gap between two polariton branches becomes
of the order of magnitude of $\sqrt{\Gamma\Omega_0}\gg \Gamma$,
where $\Omega_0$ is the frequency of the exciton transition.
These, so called Bragg MQW structures were quite intensively
studied both theoretically
\cite{Keldysh,IvchenkoSpectrum,Vladimirova,DLPRBSpectrum} and
experimentally.\cite{Hubner1996,Merle,Haas,Hayes,Khitrova} In the
case of structures with a small number of periods, the
reconstruction of the optical spectra in Bragg MQW's can be
described as a concentration of the oscillator strengths of all
oscillators in one super-radiant mode, while all other modes
become dark.\cite{IvchenkoSpectrum}

A presence of the large band gap in the spectrum of Bragg MQW
polaritons invites attempts to introduce a defect into a structure
in order to create local polariton states with frequencies in the
band gap.  This would affect the rate of  spontaneous emission as
well as other optical characteristics of the system. Such an
opportunity was first considered in
Ref.~\onlinecite{CitrinDefect}, and was studied in detail in
Refs.~\onlinecite{DefectMQWOL,DefectMQW}. It was shown
\cite{DefectMQW} that by introducing different types of defects
one can obtain optical spectra of a variety of shapes, which can
be pre-engineered through the choice of the type of a defect, its
position in the structure, the number of defects, etc. This fact
makes these systems of interest for optoelectronic application.
However, in order to be able to predict the optical spectra of
realistic structures, any theoretical calculations must deal with
the problem of inhomogeneous broadening, which is always present
in these systems due to unavoidable structural disorder present in
quantum wells. Calculations of
Refs.~\onlinecite{DefectMQWOL,DefectMQW} included inhomogeneous
broadening  using the linear dispersion
theory.\cite{KhitrovaLDT,LDT} The linear dispersion theory treats
inhomogeneous broadening as a simple addition to homogeneous
broadening, ignoring therefore, effects of the motional narrowing,
which have been very well studied in periodic multiple quantum
wells.\cite{Whittaker,Whittaker2,Kavokin1998} By doing so, the
linear dispersion theory grossly overestimates the negative
influence of inhomogeneous broadening, thereby giving
unrealistically pessimistic predictions. The latter fact was
verified in Ref.~\onlinecite{DeychNanotechnology}, where  effects
of the vertical disorder on the defect-induced features of the
spectra were considered numerically.\footnote{By vertical disorder
one understands random fluctuations of the exciton frequency from
one well to another, while the fluctuations of the frequency along
the plane of a single well are referred to as  horizontal
disorder.}

More accurately inhomogeneous broadening can be studied with the
help of the effective medium approximation, which was originally
introduced in Ref.~\onlinecite{KavokinGeneral} on the basis of
some qualitative arguments. In this approximation, a random
susceptibility of a single quantum well is replaced by its value
averaged over an ensemble of exciton frequencies, and therefore,
one only needs to take into account the horizontal disorder, since
on average all wells in the structure are assumed identical. This
approach was shown to agree well with experimental results for
both CW and time-resolved spectra,\cite{KavokinAssymetry} but so
far it has not been justified theoretically.  In this paper, we
derive this approximation from the Maxwell equations with a
non-local exciton susceptibility and demonstrate explicitly the
physical meaning of this approximation and the regions of its
applicability.

The main objective of this paper is to study the effects of the
inhomogeneous broadening in Bragg MQW structures with defects.
Here we consider only one structure, namely a GaAlAs/GaAs
structure with one of the wells replaced by a well with a
different exciton frequency.  A similar defect was considered in
Ref.~\onlinecite{DefectMQW}, where it was called an
$\Omega$-defect; we retain this terminology in this paper. The
consideration in Ref.~\onlinecite{DefectMQW} was focused mostly on
the concept of local polariton states arising in infinite Bragg
MQW structures, and on the effect of the resonant light tunnelling
arising due to these local states in long ideal Bragg MQW's. In
this paper, we  consider defect-induced modification of the
optical spectra in more realistic systems.  We take into
consideration the inhomogeneous broadening and concentrate on
structures with the number of periods readily available with
current growth technologies. Keeping in mind the potential of
these structures for applications, we study under which conditions
the defect-induced resonant features of the spectra of realistic
structures can be observed experimentally. We develop a general
understanding of the spectral properties of Bragg MQW's in the
presence of defects and disorder required for identifying
structures most interesting from the experimental and application
points of view. We also obtain a simplified analytical description
for the main features of the spectra, valid in some limiting
cases. This description is complemented by a detailed numerical
analysis. One of the surprising results is that the features
associated with the resonant tunnelling via the local polariton
state survive in the presence of disorder for much shorter lengths
of the structure than was originally
expected.\cite{DefectMQWOL,DefectMQW} This makes experimental
observation and application of these effects significantly more
attractive.

We also suggest a method of determining experimentally parameters
of homogeneous and inhomogeneous broadening of a defect well in
our structures from a single CW reflection spectrum. Such an
opportunity seems to be quite exciting from the experimental point
of view, since currently the separation of  homogeneous and
inhomogeneous broadenings requires the use of complicated
time-resolved spectroscopic techniques. Attempts to develop a
method for an independent extraction of the parameters of
homogeneous and inhomogeneous broadenings from spectra of periodic
MQW structures were undertaken in
Refs.~\onlinecite{KavokinGeneral,KavokinFourier}, but they  also
required either a time resolved spectroscopy or a not very
reliable Fourier transformation of the original CW spectra.

\section{Reflection spectrum of a MQW structure}

Within the framework of the linear nonlocal response theory,
propagation of an electromagnetic wave in a multiple quantum well
structure is governed by the Maxwell equations
\begin{equation}\label{eq:Maxwell}
 \nabla \times (\nabla \times \mathbf{E}) =
  \frac{\omega^2}{c^2}\left(\epsilon_\infty \mathbf{E} +
  4 \pi \mathbf{P}_{ext}\right),
\end{equation}
where $\epsilon_\infty$ is the background dielectric constant
assumed to be the same along the structure, $P_{ext}$ is the
excitonic contribution to the polarization defined by
\begin{equation}\label{eq:polarization}
 \mathbf{P}_{ext}(z) = \int \tilde\chi(\omega, z, z')
 \mathbf{E}(z')d z',
\end{equation}
and the susceptibility $\tilde\chi$ is
\begin{equation}\label{eq:susceptibility}
  \tilde\chi(\omega, z, z') = \tilde\chi(\omega) \Phi(z)\Phi(z'),
\end{equation}
where $\Phi(z)$ is the exciton envelope function and $z$ is the
growth direction. Considering only the 1-s heavy hole exciton
states and neglecting the in-plane dispersion of excitons, the
susceptibility can be written as\cite{Andreanireview}
\begin{equation}\label{eq:susc_frequency}
 \tilde\chi(\omega) =\frac{\alpha}{\omega_0 - \omega - i\gamma},
\end{equation}
where $\omega_0$ is the exciton resonance frequency, $\gamma$ is
the exciton relaxation rate due to inelastic processes, $\alpha =
\epsilon_\infty \omega_{LT} \pi a_B^3 \omega_0^2/c^2$,
$\omega_{LT}$ is the exciton longitudinal-transverse splitting and
$a_B$ is the bulk exciton Bohr radius.

The reflection spectrum of MQW structures is effectively described
by the transfer matrix method. For waves incident in the growth
direction of the structure, the transfer matrix describing
propagation of light across a single quantum well in the basis of
incident and reflected waves is \cite{IvchenkoSpectrum}
\begin{equation}\label{eq:transfer_matrix}
 T = \frac{1}{t}\begin{pmatrix}
    t^2 - r^2 & r \\
    -r & t
  \end{pmatrix},
\end{equation}
where $r$ and $t$ are the reflection and transmission coefficients
of a single quantum well,
\begin{equation}\label{eq:r_and_t}
 r = \frac{e^{i \phi}i\chi}{1-i\chi}, \qquad t = \frac{e^{i \phi}}{1-i\chi},
\end{equation}
$\phi = kd$, $k=\sqrt{\epsilon_{\infty}}\omega/c$ is the wave
number of the electromagnetic wave, $d$ is the period of the MQW
structure (the sum of widths of the barrier and the quantum well),
$\chi = \Gamma_0/(\omega_0 - \omega - i\gamma),$ $\Gamma_0$ is the
effective radiative rate
\begin{equation}\label{eq:Gamma_0_def}
  \Gamma_0 = \frac{\alpha}{2 k}\[\int dz \Phi(z)\cos k z\]^2,
\end{equation}
and we neglect the radiative shift of the exciton resonance
frequency.

The parameter $\gamma$ in the exciton susceptibility,
Eq.~(\ref{eq:susc_frequency}), introduces the nonradiative
homogeneous broadening due to inelastic dephasing of excitons.
Inhomogeneous broadening results from fluctuations of the exciton
transition frequency $\omega_0$ in the plane of a well caused by,
for example, imperfections of the interface between the well and
the barrier layers and/or presence of impurities. In general,
fluctuations of the exciton frequency are  described by the
$n$-point distribution functions,
$f(\omega_0(\rho_1),\omega_0(\rho_2),\ldots)$, where $\rho_i$ is a
coordinate of a point in the plane of the quantum well. However,
in the effective medium
approximation,\cite{KavokinGeneral,AgranovichI,Whittaker} adopted
in this paper, only a simple one-point function, $f(\omega_0)$, is
needed. This function is used to define an average susceptibility,
neglecting a possible dependence of the exciton envelop function
on the energy, \cite{KavokinGeneral,KavokinSpectrum,Savarona}
\begin{equation}\label{eq:inhom_susc}
  \tilde\chi = \int d\omega_0 \chi(\omega) f(\omega_0).
\end{equation}
In this approach, the inhomogeneous broadening is characterized by
the variance of the distribution function, $\Delta$.
As we already mentioned in the Introduction, the effective medium
approximation has never been actually derived, and therefore, its
theoretical status and applicability remains unclear. In the
Appendix we demonstrate how this approximation can be obtained
from a general theory of interaction between light and excitons in
a disordered quantum well, and explain why it gives such an
accurate description of the optical properties of quantum wells.

\begin{figure}
  \includegraphics[width=2.4in,angle=-90]{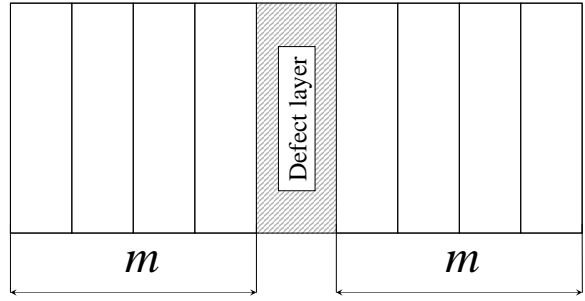}\\
  \caption{A MQW structure with a defect.}\label{fig:structure}
\end{figure}

The function $\tilde\chi$ replaces $\chi$ in
Eq.~(\ref{eq:r_and_t}) and determines a single-well transfer
matrix of a broadened well,\cite{KavokinGeneral} which is then
used to construct the transfer matrix describing the propagation
of light through the entire Bragg MQW structure. The structure
considered in this paper, a Bragg MQW with an $\Omega$-defect,
consists of $N = 2m + 1$ quantum well-barrier layers which are all
identical except for one, at the center, where the quantum well
has a different frequency of the exciton resonance (Fig.~%
\ref{fig:structure}). Such a defect can be produced either by
changing the concentration of $Al$ in the barriers surrounding the
central well,\cite{ALdependence_book,ALdependence} or the width of
the well itself\cite{widthdependence} during growth. In principle,
both these changes will also affect the optical width of the
defect layers, either because of the change in the background
dielectric constant, or due to the change of the geometrical
thickness of the well. However, in both cases this effect is
negligible, and we deal here with the case of a pure
$\Omega$-defect.

The total transfer matrix through the MQW structure consists of
the product
\begin{equation}\label{eq:total_transfer_matrix}
  M = T_h \ldots T_h T_d T_h \ldots T_h,
\end{equation}
where $T_h$ and $T_d$ are the transfer matrices through the host
and defect layers, respectively, described by the reflection and
transmission coefficients $r_{h,d}$ and $t_{h,d}$. Substitution of
the explicit expressions for $T_h$ and $T_d$ leads to a compact
expression for the total transfer matrix in the basis of
eigenvectors of the host transfer matrix $T_h$
\begin{equation}\label{eq:M_explicit}
  M =
  \begin{pmatrix}
   e^{-\Lambda} M_- & a_+A \\
    -a_- A & e^{\Lambda}M_+
  \end{pmatrix} ,
\end{equation}
where $\Lambda = N\lambda_h$,
\begin{equation}\label{eq:Mpm_defs}
  M_\pm = e^{\pm (\lambda_d - \lambda_h)}
   \pm \frac{2 e^{\mp\lambda_h}}{\sinh \lambda_h} \sinh^2 \frac 1 2
(\lambda_d -  \lambda_h),
\end{equation}
and
\begin{equation}\label{eq:A_def}
  A = \frac{\sin \phi}{\sinh \lambda_h}\left(\chi_d - \chi_h\right).
\end{equation}

Here we introduced $a_{\pm}$, non-unit components of the
eigenvectors of $T_h$,
\begin{equation}\label{eq:ab_eqs}
  a_\pm = \frac{1-e^{\pm\lambda_h}t_h}{r_h},
\end{equation}
and $\lambda_{h,d}$ are the eigenvalues of the host and defect
quantum well's transfer matrices obeying the dispersion law in a
periodic quantum well
superlattice\cite{Keldysh,IvchenkoSpectrum,CitrinSpectrum}:
 \begin{equation}\label{eq:lambdas_eqs}
  \cosh \lambda_{h,d} = \frac 1 2 \mathrm{Tr}\, T_{h,d} =
  \cos \phi - \chi_{h,d}\sin \phi.
\end{equation}
In the case of an ideal system without homogeneous or
inhomogeneous broadenings, this equation describes the band
structure of the electromagnetic spectrum of MQW's, consisting of
a number of bands separated by forbidden band gaps, defined as
frequency regions where $\mathrm{Re}\,\lambda_h \ne0$. This real
part describes the exponential decrease of the amplitude of an
incident wave, and its inverse  gives the value of the respective
penetration length. In the allowed bands, $\lambda_h$ is purely
imaginary, and its imaginary part is the Bloch wave vector of the
respective excitation. There are two types of the excitations
here. In the vicinity of the exciton frequency $\omega_h$, there
exist two polariton branches separated by a polariton band
gap.\cite{DLPRBSpectrum} There are also pure photonic bands with
the band boundaries at the geometrical resonances, $\phi(\omega_r)
= n\pi$ ($n=1,2\ldots$). The size of the polariton gap strongly
depends on the relation between $\omega_h$ and $\omega_r$, and
reaches the maximum value when they coincide, i.e. when
$\phi(\omega_h) = \pi$ (the Bragg structure). For these
frequencies, it is convenient to extract the imaginary part from
$\lambda_h$ and to present it in the form
\begin{equation}\label{eq:l_h_repr}
  \lambda_h = \kappa_h + i \pi.
\end{equation}
If the coupling parameter $\Gamma_0$ is small, $\kappa_h$ in the
gap can be approximated by the following expression, which is
obtained by expanding Eq.~(\ref{eq:lambdas_eqs}) near the
resonance frequency
\begin{equation}\label{eq:k_h_appr}
  \kappa_h =  \sqrt{\pi q(-2\chi_h - \pi q)},
\end{equation}
where $q$ is the detuning from the Bragg resonance
\begin{equation}\label{eq:q_def}
  q = \frac{\omega - \omega_h}{\omega_h},
\end{equation}
Taking into account the form of the susceptibility in an ideal
system, we obtain
\begin{equation}\label{eq:kappa_h_gap}
  \kappa_h = \pi\sqrt{Q_G^2 - q^2},
\end{equation}
where
\begin{equation}\label{eq:gap_size}
  Q_G = \sqrt{\frac{2\Gamma_0}{\pi \omega_h}}
\end{equation}
determines the boundary of the forbidden gap\cite{DLPRBSpectrum}
as a point where $\kappa_h$ vanishes. The frequency, which
corresponds to this boundary determines the width of the gap as
$\sqrt{2\Gamma\omega_h/\pi}$. This value is significantly greater
than the respective width in the off-Bragg case, which is
proportional to $\Gamma\ll\omega_h$. In the presence of
homogeneous and inhomogeneous broadenings, the notion of the band
gap becomes ill defined, because $\kappa_h$ does not vanish
anywhere. At the gap boundary, for instance, it becomes complex
valued
\begin{equation}\label{eq:kappa_at_the_edge}
  \kappa_h(\omega_G) = (1 + i) \pi \sqrt{\frac{\gamma
  Q_G}{2\omega_h}}.
\end{equation}
 Nevertheless, if
$\gamma\ll \sqrt{\Gamma\omega_h}$, which is the case in realistic
systems, the optical spectra retain most of their properties
specific for the gap region, and this concept provides a useful
physical framework for discussing optical properties of MQW
structures.


\begin{figure}
  \includegraphics[width=3in,angle=-90]{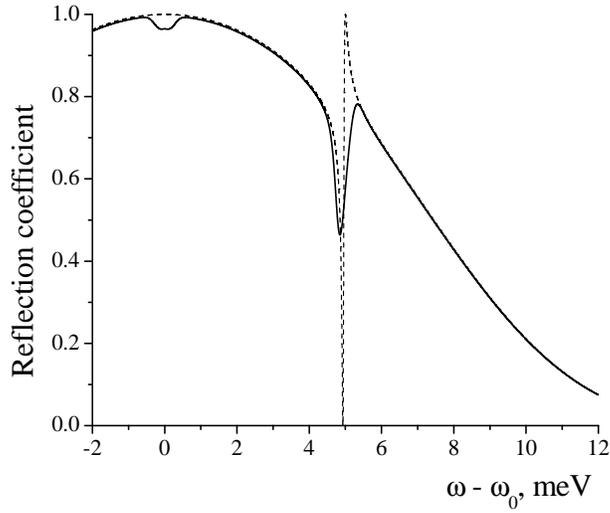}
  \caption{A typical dependence of the reflection coefficient
 of a MQW structure with an embedded defect well in a neighborhood of
 the exciton frequency of the defect well.
 The dotted line shows the reflection for a lossless system, and the solid
 line corresponds to a broadened system (parameters are taken for
 GaAs/AlGaAs).}
\label{fig:typical_resonance}
\end{figure}

Once the transfer matrix, $M$, for the entire MQW structure is
known the reflection and transmission coefficients can be
expressed in terms of its elements, $m_{ij}$, as
\begin{equation}\label{eq:total_r_and_t}
\begin{split}
  r_{MQW} = \frac{-m_{11} + m_{12} - m_{21} + m_{22}}
      {a_-\left( m_{12} + m_{22} \right) - a_+\left( m_{11} + m_{21} \right)},
\\
  t_{MQW} = \frac{a_- - a_+}
      {a_-\left(m_{12} + m_{22}\right) - a_+\left( m_{11} + m_{21} \right)}.
\end{split}
\end{equation}
A defect inserted into the structure leads to a modification of
the reflection spectrum of the MQW in the vicinity of the exciton
frequency, $\omega_d$,  of the defect well. We are interested in
the situation when $\omega_d$ lies within the polariton band gap
of the ideal host structure, because in this case the defect
produces the most prominent changes in the
spectra. 
A typical form of such a modification in broadened systems is
shown in Fig.~(\ref{fig:typical_resonance}), and is characterized
by the presence of a closely positioned minimum and maximum. In an
ideal system, this form of reflection would correspond to a
Fano-like resonance in the transmission with the latter swinging
from zero to unity over a narrow frequency
interval.\cite{DefectMQW} In the presence of absorbtion and
inhomogeneous broadening, the resonance behavior of the
transmission is smeared out, while the resonance in reflection, as
we can see, survives.

In order to analyze the form of the reflection spectra, we
represent the reflection coefficient in the form
\begin{equation}\label{eq:r_add_def}
  r_{MQW} = \frac{r_0}{1 - r_{add}},
\end{equation}
where
\begin{equation}\label{eq:r_pure}
  r_0 = \frac{2 \sinh (\Lambda)}{a_- e^{\Lambda}
    - a_+ e^{-\Lambda}} = -
    \frac{\chi_h}{\alpha + i \coth(\Lambda)\sinh\lambda_h}
\end{equation}
is the reflection coefficient of a pure MQW structure (without a
defect) with the length $N$, $\alpha = \sin\phi + \chi_h
\cos\phi$, and $r_{add}$ introduces the modification of the
reflection caused by the defect
\begin{eqnarray}\label{eq:r_add_details}
  r_{add} = (\chi_d - \chi_h)\sin \phi\, \nonumber
    \frac{\sinh\lambda_h + i\chi_h\sinh\Lambda}{\chi_h\cosh\Lambda -
    \alpha}\\
  \times \frac{1}{A (\alpha + \chi_h\cosh \Lambda) - \chi_h\sinh\Lambda}.
\end{eqnarray}
This expression allows for making some general conclusions
regarding the effects of the defect and broadenings on the
reflection spectrum. First of all, it should be noted that
regardless of the value of the defect frequency $\omega_d$,
$r_{add}$ vanishes at frequencies $\omega=\omega_h$ because of the
phase factor $\sin\phi \approx -\pi q$.

Thus, in order to achieve a significant modification of the
spectrum, it is necessary to choose $\omega_d$ as far away from
$\omega_h$ as possible. In this case, however, we can immediately
conclude that the broadening of the host wells does not
significantly affect the defect-induced features of the reflection
spectrum.

Indeed, the broadenings enter into  Eq.~%
(\ref{eq:r_add_details}) through the susceptibility $\chi_h$
defined by Eq.~(\ref{eq:inhom_susc}). Let us rewrite this
definition in the form
\begin{equation}\label{eq:susc_host_qual}
  \chi_h = \int d \nu f(\nu)\frac{\tilde\Gamma}{\nu - q -i\tilde\gamma},
\end{equation}
where we have introduced $\tilde\Gamma = \Gamma_0/\omega_h$,
$\tilde\gamma = \gamma/\omega_h$, $\nu = (\omega_0 -
\omega_h)/\omega_h$, and $\omega_h$ is the mean value of the
exciton frequency. If the function $f(\nu)$ falls off with
increasing $\nu$ fast enough, so that all its moments exist, 
we can approximate the integral  for the frequencies $q/\tilde\Delta,\, q/\tilde\gamma \gg 1$ as
\begin{equation}\label{eq:susc_host_app}
 \chi_h \approx \tilde\chi_h \int d \nu f(\nu)\left(1 + \frac{\nu}{q +i\tilde\gamma} + \ldots\right),
\end{equation}
where
\begin{equation}\label{eq:susc_host_bare}
  \tilde\chi_h = \frac{\Gamma_0}{\omega_h - \omega - i\gamma}
\end{equation}
is the susceptibility in the absence of the inhomogeneous
broadening. Noting that now integration of each term in the
parentheses gives an appropriate central moment of $\nu$ we obtain
\begin{equation}\label{eq:susc_host_app2}
 \chi_h \approx \tilde\chi_h \left(1 + \frac{\tilde\Delta^2}{q^2} + \ldots\right).
\end{equation}
Therefore, the corrections due to the inhomogeneous broadenings
become small for frequencies, which are farther away from the
central frequency than the inhomogeneous width $\Delta$. Since the
width of the polariton gap in Bragg MQW structures is
significantly greater than the typical value of the inhomogeneous
width, we can choose such a position of the defect frequency,
$\omega_d$, which is  far enough from $\omega_h$, and at the same
time remains within the gap frequency region.

Significant diminishing of the effects due to disorder in optical
spectra of periodic MQW for frequencies away from the resonance
exciton frequency was obtained theoretically in
Ref.~\onlinecite{KavokinGeneral} and observed experimentally in
Ref.~\onlinecite{InhomFar}. This can also be seen in
Fig.~\ref{fig:typical_resonance}, where deviations from the
spectra of an ideal structure (dashed line) caused by  disorder in
the host wells are significant only in the vicinity of $\omega_0$.
The modification of the defect induced features of the spectra,
which takes place in the vicinity of $\omega_d$ are caused by
broadenings of the excitons in the defect layer, and this is the
only broadening, which has to be taken into consideration in this
situation.

\section{Deep and shallow defects}

While Eq.~(\ref{eq:r_add_details}) is suitable for a general
discussion and numerical calculations, it is too cumbersome for a
detailed analytical analysis. For such purposes we consider the
reflection coefficient for two limiting cases when the expression
for $r_{add}$ can be significantly simplified. The first limit
corresponds to the situation when the penetration length of the
electromagnetic wave at the frequency of the defect is much
smaller than the length of the structure (we call it a deep
defect), and the second is realized in the opposite case, when the
system is much shorter than the penetration length (shallow
defect).

\subsection{Deep defect}

When $\omega_d$ lies so far away from both $\omega_h$ and the
boundary of the band gap that the penetration length of the
electromagnetic wave is much smaller than the length of the
structure [$\mathrm{Re}(\Lambda) \gg 1$], two different
approximations are possible. When the homogeneous broadening is
small enough such that
\begin{equation}\label{eq:hom_limit}
  \gamma < 8\pi \omega_h \sqrt{Q_G^2 -
  \delta^2}\,\frac{e^{-2\Lambda}\delta^3}{Q_G^2},
\end{equation}
where $\delta = (\omega_d - \omega_h)/\omega_h$, the systems is
close to ideal, and the results of Ref.~\onlinecite{DefectMQW}
remain valid. This inequality, however, becomes invalid for long
systems, and then the exponentially small non-resonant terms in
Eq.~(\ref{eq:r_add_details}) can be neglected. The reflection in
the vicinity of $\omega_d$ in this case can be presented in the
form
\begin{equation}\label{eq:ref_deep_Lorentz}
r = r_0\frac{\Omega_d - \Gamma_0 D_d}{
   \Omega_d - \Gamma_0 D_d - 2ie^{-\Lambda}\delta^2\omega_h},
\end{equation}
where $D_{d,h} = 1/\chi_{d,h}$ and
\begin{equation}\label{eq:r_add_deep}
 r_0 = \frac{1}{1 + D_h \left[ \pi q + i \kappa_h
    \left(1 + 2 e^{-2\Lambda}\right)\right]}
\end{equation}
is the approximation for the reflection coefficient of the
structure without the defect for frequencies deeply inside the
forbidden gap. We keep the term $\exp(-2\Lambda)$ in this
expression in order to preserve the correct dependence of the
reflection coefficient of the pure structure on its length. The
frequency $\Omega_d$,
\begin{equation}\label{eq:shift_deep}
 \Omega_d = \pi\Gamma_0\, \frac{\delta}{\kappa_h},
\end{equation}
describes the shift of the position of the reflection resonance
from the initial defect frequency $\omega_d$. This shift is an
important property of our structure, which takes place in both
ideal and broadened systems; Eq.~(\ref{eq:shift_deep}) is a
generalization to the systems with inhomogeneous broadening of the
result obtained in Ref.~\onlinecite{DefectMQW}.

Deriving Eq.~(\ref{eq:ref_deep_Lorentz}), in addition to the
assumption about the relation between the length of the structure
and the penetration length, we also assumed that the exciton
frequency of the defect well lies far enough from the frequency of
the host wells, and neglected the contribution of the host
susceptibility $\chi_h$ into the terms proportional to $\chi_d -
\chi_h$. We also dropped a frequency dependence of the
non-resonant terms.

Eq. (\ref{eq:ref_deep_Lorentz}) shows that when the defect well
exciton frequency lies deeply inside the forbidden gap the effect
of the defect on the reflection spectrum of the system
exponentially decreases when the length of the MQW structure
increases. This behavior is strikingly different from that of the
respective ideal systems, where resonant tunnelling results in the
transmission equal to unity at the resonance regardless of the
length of the system. One can see that homogeneous broadening
severely suppresses this effect, as was anticipated in
Ref.~\onlinecite{DefectMQW}.

If the shift, $\Omega_d$, of the resonance frequency  from
$\omega_d$ is large enough, so that $\omega_r$ is well separated
from $\omega_d$, the effects of the inhomogeneous broadening can
be neglected. In this case, we can derive a simple approximate
expression for the reflection coefficient in the vicinity of
$\omega_r$. The condition $\Omega_d \gg \Delta$ can, in principle,
be fulfilled because $\kappa_h(\omega_d)$ decreases when the
frequency goes to the edge of the stop band where $\kappa_h$ is
determined by Eq.~(\ref{eq:kappa_at_the_edge}) and for GaAs/AlGaAs
MQW structures with $\omega_h = 1.49 \,\mathrm{eV}$, $\Gamma =
67\, \mathrm{\mu eV}$, $\gamma = 12.6\, \mathrm{\mu eV}$ and
$\Delta = 290\, \mathrm{\mu eV}$ we obtain
$\mathrm{Re}[\Omega(\omega_h + \omega_G )]/\Delta \approx 6.3$.

In this case, in the vicinity of the resonance frequency we can
approximate the susceptibility $\chi_d$ by
\begin{equation}\label{eq:deep_chi_d_no_D}
  \chi_d = \frac{\Gamma_0}{\omega_d - \omega - i\gamma}
\end{equation}
and obtain that the resonance has a form of the Lorentz-type dip
on the dependence of the reflection spectrum positioned at $q =
\Omega_d$ with the depth, $H$, and the width, $W$, defined by
expressions
\begin{equation}\label{eq:deep_h_no_D}
  H = |r_0|^2\frac{1 + \gamma e^\Lambda/\omega_h\delta^2}{
    \left(1 + \gamma e^\Lambda/2\omega_h\delta^2\right)^2},
%
\end{equation}
\begin{equation}\label{eq:w_no_D}
W = \gamma + e^{-\Lambda}\delta^2\omega_h.
\end{equation}
It should be noted, however, that while formally this
approximation is valid even when $\omega_d$ is close to the edge
of the forbidden gap, the deep defect approximation requires that
$\Lambda\gg 1$. For $GaAs/AlGaAs$ structures this means that $N >
1/\mathrm{Re}(\kappa_h) \sim 2000$. The structures of this length
are beyond current technological capabilities, so this case
presents mostly theoretical interest.

In the opposite situation, when the frequency shift is small
($\Omega_d < \Delta$), i.e. when $\omega_d$ is not too close to
the edge of the gap, the inhomogeneous broadening becomes
important, and in order to estimate its contribution we use a
Gaussian distribution of the exciton resonance frequencies in Eq.~%
(\ref{eq:inhom_susc}):
\begin{equation}\label{eq:inhom_susc_Gaussian}
  \chi_{d} = \frac{\Gamma_0}{\Delta \sqrt{\pi}}
  \int_{-\infty}^\infty d\omega_0\, \frac{e^{-(\omega_0 - \omega_d)^2/\Delta^2}}{\omega_0 - \omega -
  i\gamma}.
\end{equation}
Using the function $w(\mu) = e^{-\mu^2}\mathrm{erfc}(-i\mu)$, the
integral can be written as
\begin{equation}
\chi = i\Gamma_0 w(\mu)\sqrt{\pi}/\Delta,
\end{equation}
where $\mu = (\omega - \omega_d + i\gamma)/\Delta$. The small
$\mu$ expansion\cite{Abramowitz}
\begin{equation}  \nonumber
w(\mu) \approx 1 + 2 i \mu /\sqrt{\pi}
\end{equation}
allows us to obtain:
\begin{equation}\label{eq:D_small_mu}
  D_d = \frac{2}{\Gamma_0 \pi}\left(\omega_d - \omega -
   i\tilde \gamma\right),
\end{equation}
where $\tilde\gamma$ is the effective broadening,
\begin{equation}\label{eq:gamma_renorm_inhom}
  \tilde\gamma = \gamma + \frac{\sqrt{\pi}}{2}\Delta.
\end{equation}
One can see that in this case the inhomogeneous and homogeneous
broadening combine to form a single broadening parameter
$\tilde\gamma$, as it is assumed in the linear dispersion theory.
The resonance on the reflection curve also has, in this case, a
Lorentz-type dip centered at
\begin{equation}\label{eq:deep_resonance_point}
  q = \frac{\pi \Gamma_0
  \delta}{2\omega_h\sqrt{\omega_G^2-\delta^2}},
\end{equation}
with the depth and the width, respectively, equal to
\begin{equation}\label{eq:deep_far}
  H = |r_0|^2 \frac{2\pi e^{-\Lambda}\delta^2 \omega_h}{\tilde\gamma},
  \qquad
  W = \tilde\gamma + \pi e^{-\Lambda}\delta^2\omega_h.
\end{equation}
Because of the  inhomogeneous broadening contribution, the
effective parameter $\tilde\gamma$ becomes so large that $\pi
e^{-\Lambda} \delta^2 \ll \tilde\gamma/\omega_h$, and the
defect-induced reflection resonance becomes rather weak compared
to the case considered previously.


\subsection{Shallow defect}

From the experimental point of view, a more attractive situation
arises when the length of the structure is smaller than the
penetration length. This situation, which can be called the case
of a shallow defect, can be described within the same
approximations as used in the previous section, with an obvious
exception of the treatment of terms proportional to $e^\Lambda$.
Here we can expand the exponential function in terms of the powers
of $\Lambda$. Finally we arrive at the following expression, which
describes the defect-induced modification of the reflection
spectra:
\begin{equation}\label{eq:r_edge_add}
 r = \frac{i\bar\Gamma}{\omega_h - \omega + i(\gamma +
 \bar\Gamma)}\,
  \frac{\Omega_s -\Gamma_0 D_d}{i \Gamma_0 - \Gamma_0 D_d},
\end{equation}
where $\bar\Gamma$ is the  radiative width of the pure Bragg MQW
structure, which is $N$-fold enhanced because of the formation of
a superradiant mode\cite{IvchenkoMQW}
\begin{equation}\label{eq:Gamma_bar}
 \bar\Gamma = \frac{\Gamma_0 N}{1- i \pi q N}.
\end{equation}
This expression coincides with the results of
Ref.~\onlinecite{IvchenkoMQW} with the exception of the term
proportional to $qN$, which was neglected in the previous papers.
We keep this term to be able to consider the case when the
detuning from the resonance point $\omega_h$ is not very
small\cite{IvchenkoContrast}.

The distinctive feature of the shallow defect is that the
reflection resonance does not have a Lorentz-like shape, which is
typical for the deep defect when the resonant tunnelling is
suppressed. Instead, we have a reflection spectrum with a minimum
at $\omega_-$ and a maximum at $\omega_+$. This is a surprising
result, because the Fano-like behavior associated with the
resonant tunnelling is restored even though the length of the
system is too short for effective tunnelling to take place. It is
convenient to describe the positions of these frequencies relative
to the modified defect frequency,
\begin{equation}
\tilde\omega_d=\omega_d-\Omega_s,
\end{equation}
where $\Omega_s$ is defined as
\begin{equation}\label{eq:Omega_s}
  \Omega_s = \frac{\omega_d - \omega_h}{N}.
\end{equation}
The positions of the extrema are shifted from $\tilde\omega_d$:
$\omega_-$ toward the center of the gap, and $\omega_+$ in the
opposite direction. As a result, $\omega_-$ is well separated from
$\omega_d$, while $\omega_+$ always lies in the close vicinity of
the defect frequency.

For \textit{short} systems, $\Omega_s$ can become larger than
$\Delta$; for example, in GaAs/AlGaAs MQW structures, the
condition $\Omega_s \gg \Delta$ is fulfilled when $N \lesssim N_0
=10$. In this case, an approximate analytical description of the
spectrum is possible again. However, since the maximum and the
minimum of the spectra lie at significantly different distances
from $\omega_d$, the description of these two spectral regions
would require different approximations. The maximum of the
reflectivity takes place close to the defect frequency, and
therefore the inhomogeneous broadening near the maximum has to be
taken into account. At the same time, $\omega_-
-\omega_d\gg\Delta$, and the inhomogeneous broadening in this
frequency region can be neglected. Thus, we can approximate $D_d$
by Eq.~(\ref{eq:deep_chi_d_no_D}) in the vicinity of $\omega_-$
and by Eq.~(\ref{eq:D_small_mu}) near $\omega_+$. Using these
approximations we find that the minimum and the maximum of the
reflection coefficient are at the frequencies
\begin{equation}\label{eq:shallow_minimum}
 \omega_- = \omega_d - \Omega_s - \frac{\gamma^2}{\Omega_s}
\end{equation}
and
\begin{equation}\label{eq:shallow_maximum}
 \omega_+ = \omega_d + \frac{1}{\pi}\left(\tilde\Omega_s -
 \Omega_s\right) + \frac{\Gamma_0 \tilde\gamma}{\tilde\Omega_s\Omega_s}
 \left(\tilde\Omega_s + \Omega_s\right)
\end{equation}
respectively, where $\tilde\Omega_s = \sqrt{\Omega_s^2 +
4\tilde\gamma^2}$ and $\tilde\gamma$ is the effective broadening
defined by Eq.~(\ref{eq:gamma_renorm_inhom}). The values of the
reflection at these points are
\begin{eqnarray}\label{eq:shallow_R_min_max}
 R_{min} = \frac{|\bar\Gamma|^2\gamma^2 N^4}{(\omega_d -
  \omega_h)^4(N - 1)^2} , \nonumber \\
 R_{max} = \frac{|\bar\Gamma|^2 (\tilde\Omega_s + \Omega_s)^2}{
    (\omega_+ - \omega_h)^2 (2\Gamma_0 + \pi\tilde\gamma)^2} .
\end{eqnarray}

\begin{figure}
  \includegraphics[width=3in,angle=-90]{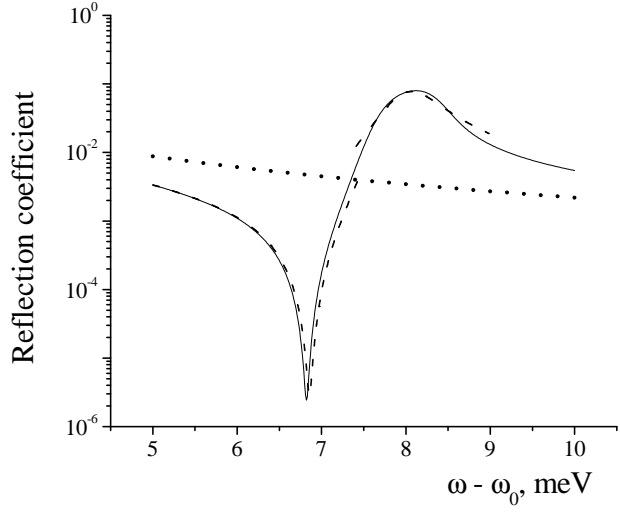}
  \caption{Reflection coefficient near the exciton frequency of
  the shallow defect (solid line) for $N = 7$. The dashed lines depict
  approximation using different expressions for the defect
  quantum well susceptibility at the vicinities of the extrema: near the
  minimum the inhomogeneous broadening is neglected, while in the
  vicinity of the maximum it is accounted for as a renormalization of homogeneous
  broadening [Eq.~(\ref{eq:gamma_renorm_inhom})]. For reference,
  the reflection coefficient of a pure MQW structure without
  a defect is shown (dotted line).}
\label{fig:shallow_plus_minus}
\end{figure}

The exact and approximate forms of the reflectivity are compared
in Fig.~\ref{fig:shallow_plus_minus}. One can see that these
approximations give a satisfactory description of the reflectivity
in the vicinities of the extrema for short systems.

\begin{figure}
  \includegraphics[width=3in,angle=-90]{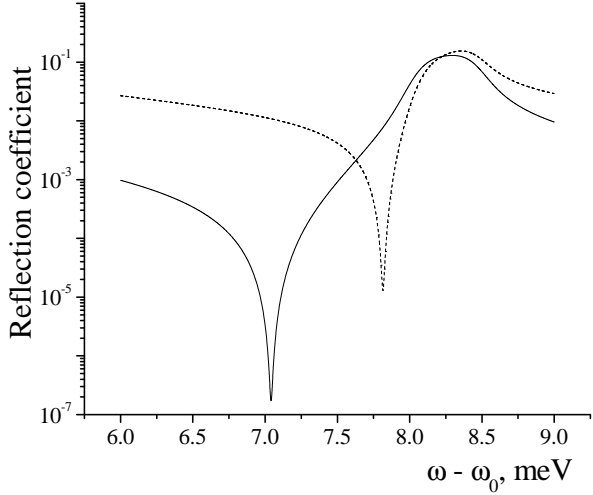}
  \caption{Dependence of the reflection coefficient on the
frequency in the neighborhood of the exciton frequency of the
defect well for different lengths of the MQW structure (solid line
$N = 3$, dashed line $N = 5$). The great difference of minimal
reflections results from the shift of the resonance frequency
$\Omega_s$, Eq.~(\ref{eq:Omega_s}).}
\label{fig:shallow_comparisons}
\end{figure}

The minimal value of the reflection is determined only by the
small parameter of the homogeneous broadening, $\gamma$,  and can
therefore become very small. This fact reflects the suppression of
the inhomogeneous broadening in this situation. When the length of
the system increases, $R_{min}$ grows as $N^4$, however, when $N
> N_0$, the inhomogeneous broadening starts coming into play:
$\gamma$ must be replaced with a larger effective broadening
containing a contribution from $\Delta$. This also leads to a
significant increase in $R_{min}$. This behavior is illustrated in
Fig.~\ref{fig:shallow_comparisons}, where a comparison of the
reflection coefficients for two MQW structures with different
lengths is provided. The expression for $R_{min}$,
Eq.~(\ref{eq:shallow_R_min_max}), allows one to obtain an estimate
for the parameter of the homogeneous broadening $\gamma$ by
measuring the value of $R_{min}$, since all other parameters
entering this expression are usually known. This approach to
determining $\gamma$ from reflection experiments, however, cannot
be used when the predicted values of $R_{min}$ is too small,
because other factors (such as small mismatch in the indexes of
refraction between different components of the structure) which
were neglected in our calculations can become important. At the
same time, it can be expected that for not very short systems,
satisfying the condition $N<N_0$, the parameter $\gamma$ can be
determined from the reflection spectrum.

The inhomogeneous parameter $\Delta$ enters expressions
Eq.~(\ref{eq:shallow_maximum}) for $\omega_+$ and
Eq.~(\ref{eq:shallow_R_min_max}) for $R_{max}$. Therefore, one can
determine this parameter from two independent values accessible
from the reflection spectra. The maximum value of the reflection
coefficient in this approximation depends very weakly on the
number of wells in the system, and is of the order of the
magnitude of the reflection from a single standing defect well in
the exact resonance. This result means that for a small number of
wells, $\omega_+$ lies in the spectral region, where the host
system is already almost transparent, and the presence of the host
has only a small effect on the reflection properties of the system
with the defect. At the same time, we would like to emphasize
again that the minimum of the reflection is the result of the
radiative coupling between the wells even in this case of short
systems.

\subsection{Characterization of the reflection spectra in the case of intermediate lengths}

In the previous subsections we examined special situations when
the defect can be considered as either deep or shallow. In both
cases, we were able to derive approximate analytical expressions
describing defect-induced modification of the spectra and to
obtain a qualitative understanding of how the defect affects the
reflection spectrum. In particular, it was demonstrated  that the
characteristic frequencies related to the modification of the
spectrum are shifted from the resonant defect frequency of a
single well. This shift is the result of radiative coupling
between the excitons in the defect well and the collective
excitations of the host system. One of the important consequences
of this shift is the possibility for almost complete suppression
of the effects due to inhomogeneous broadening in some spectral
intervals. In this subsection we consider systems with
intermediate lengths, when $N$ is larger than $N_0$, but is still
smaller or of the order of magnitude of the penetration length.
From the practical point of view, this case is of the greatest
interest, since this interval of lengths is still easily
accessible experimentally, and at the same time, it is expected
that for such structures the defect-induced modifications of the
spectrum become most prominent. Unfortunately, neither
approximation used in the previous subsections can be applied
here, and we have to resort to a numerical treatment.
Nevertheless, the qualitative understanding gained as a result of
the previous analytical considerations, serves as a useful guide
in analyzing and interpreting the numerical data.

As it was pointed out in the previous section, when $N$ becomes
larger than $N_0$, the position of the minimum of the reflection,
$\omega_-$ moves closer to $\omega_d$, and the inhomogeneous
broadening starts contributing to $R_{min}$. This effect can
phenomenologically be described as the emergence of an effective
broadening parameter $\gamma_{eff}(\gamma,\Delta,N)$, which is not
a simple combination of $\gamma$ and $\Delta$, but depends upon
$N$. This parameter is limited from below by $\gamma$, when the
inhomogeneous broadening is suppressed, and from above by
$\tilde\gamma$, when the contribution from $\Delta$ is largest.
Because the minimum value of the reflection is always achieved at
a point shifted with respect to $\omega_d$, generally
$\gamma_{eff}$ is always smaller than $\tilde\gamma$, and the
homogeneous broadening makes the main contribution to it even for
systems with $N>N_0$ {\it despite the fact that
$\gamma\ll\Delta$}. At the same time, the position of $\omega_+$,
which determines the width of the spectral interval affected by
the defect, depends upon $\tilde\gamma\approx\Delta$. Thus the
effect of the broadenings on the spectrum can in general be
summarized in the following way: while the width of the resonance
is determined equally by both homogeneous and inhomogeneous
broadenings, its strength depends mostly upon the homogeneous
broadening.

\begin{figure}
\includegraphics[width=3in,angle=-90]{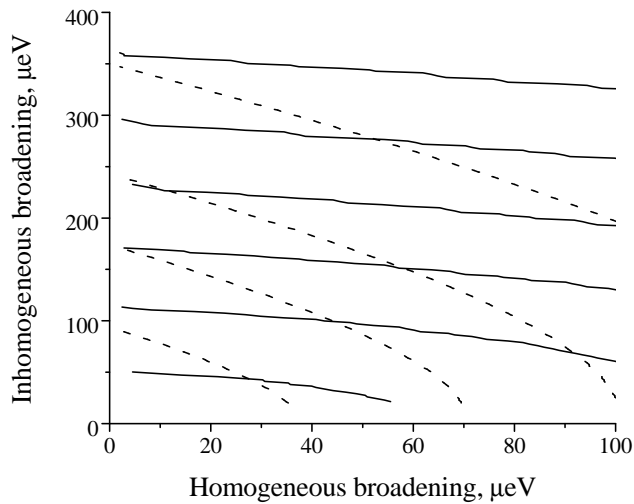}\\
  \caption{Intersections of lines of constant height (dashed lines) and
  width (solid lines) of the
  resonance allow determination of the values of the homogeneous and
  inhomogeneous broadenings.}\label{fig:constant_level}
\end{figure}

We  illustrate this conclusion quantitatively by defining the
width of the resonance, $W(\gamma,\Delta,N) = \omega_+ -
\omega_-$, as a distance between the extrema of the reflection
spectrum, and its depth, $H(\gamma,\Delta,N) = R_{max} - R_{min}$,
as the difference between the values of the reflection at these
points. In order to see how these quantities depend upon
parameters $\gamma$ and $\Delta$,  we chose several different
values of $W$ and $H$, and plot constant level lines,
$W(\gamma,\Delta,N)=W_i$, $H(\gamma,\Delta,N)=H_i$. These lines
represent values of $\gamma$ and $\Delta$ for which $W$ and $H$
remain constant (Fig.~\ref{fig:constant_level}).

The locus of  constant widths is the set of nearly straight lines
running almost parallel to the axis representing the homogeneous
broadening. Slight deviation from the straight-line behavior is
seen only for non-realistically high values of $\gamma$. Such a
behavior confirms our assertion that the width of the resonance is
determined  by an effective parameter, in which $\gamma$ and
$\Delta$ enter additively. As we see, this is true even for
systems which cannot, strictly speaking, be described by
approximations leading to Eq.~(\ref{eq:shallow_maximum}). Since
usually $\gamma\ll \Delta$, the latter makes the largest
contribution to this effective parameter, and determines the value
of $W$. The shape of the lines of constant height demonstrates
almost equal contributions from $\gamma$ and $\Delta$, which means
that the effect due to the inhomogeneous broadening is
significantly reduced as far as this feature of the spectrum is
concerned. This is also consistent with an approximate analysis
presented in the previous section of the paper.

The remarkable feature of Fig.~\ref{fig:constant_level} is that
the lines of the constant width and the constant high cross each
other at a rather acute angle and at a single value of $\gamma$
and $\Delta$ for each of the values of $W$ and $H$. This means
that one can extract both $\gamma$ and $\Delta$ from a single
reflection spectrum of the MQW structure. This is a rather
intriguing opportunity from the experimental point of view, since
presently, the only way to independently measure parameters of
homogeneous and inhomogeneous broadenings is to use complicated
time-resolved techniques.

It is clear, however, that the shape of the lines of constant $W$
and $H$ depends upon the choice of the distribution function used
for calculation of the average susceptibility of the defect well.
It is important, therefore, to check how the results depend upon
the choice of the distribution function of the exciton
frequencies. As an extreme example, one can consider the Cauchy
distribution
\begin{equation}\label{eq:}
  f(\omega_0) = \frac{\Delta}{(\omega_0 - \bar\omega)^2 +
  \Delta^2}.
\end{equation}
In this case, all the effects due to the inhomogeneous broadening
can be described by a simple renormalization, $\gamma \rightarrow
\gamma + \Delta$, and the level lines in
Fig.~(\ref{fig:constant_level}) would have the form of parallel
lines. This distribution though, hardly has any experimental
significance, while the Gaussian function has a certain
theoretical justification.\cite{Efros} However, the symmetrical
character of the normal distribution is in  obvious contradiction
with a natural asymmetry of the exciton binding energies, which
can be arbitrarily small but are bounded from above. It was
suggested in Ref.~\onlinecite{KavokinAssymetry} to take this
asymmetry into account by introducing two different variances in
the Gaussian distribution: $\Delta_-$ for frequencies below some
(most probable) frequency $\omega_c$, and $\Delta_+$ for the
frequencies above it. Accordingly, the distribution function can
be written as
\begin{equation}\label{eq:Gaussian_assymetry}
  f(\omega_0) = \frac{2}{\sqrt{\pi}(\Delta_+ + \Delta_-)}
  \begin{cases}
    e^{-\frac{(\omega_0 - \omega_c)^2}{\Delta_-}}, & \omega_0 < \omega_c,\\
    e^{-\frac{(\omega_0 - \omega_c)^2}{\Delta_+}}, & \omega_0 > \omega_c.
  \end{cases}
\end{equation}
It was shown that this choice gives a satisfactory description of
time-resolved spectra of MQW's.\cite{KavokinAssymetry} The
distribution function Eq.~(\ref{eq:Gaussian_assymetry}) can be
parameterized either by $\Delta_\pm$ and $\omega_c$ or, more
traditionally, by the mean value, $\bar\omega$, the second moment,
$\Delta$, and the parameter of asymmetry, $n$, defined as
\begin{eqnarray}\label{eq:distribution_parameters}
  \bar\omega = \frac{\Delta_+ - \Delta_-}{\sqrt{\pi}}, \quad
    n = \frac{\Delta_+}{\Delta_-}, \nonumber
  \\   %
  \Delta^2 =  \frac{\Delta_+^3 + \Delta_-^3}{\Delta_+ + \Delta_-}
    - \frac{2}{\pi}(\Delta_+ - \Delta_-)^2.
\end{eqnarray}

\begin{figure}
  \includegraphics[width=3in,angle=-90]{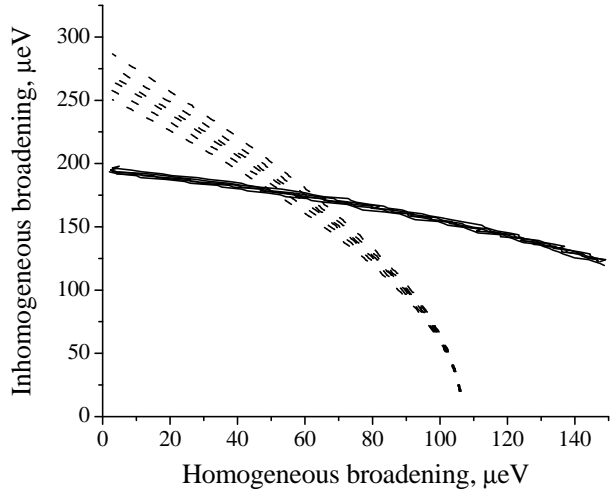}\\
  \caption{Intersections for different values of the parameter of
  asymmetry. Dashed and solids lines are lines of constant height
  and width respectively.}\label{fig:levels_asymmetry}
\end{figure}

We use the corrected distribution function,
Eq.~(\ref{eq:Gaussian_assymetry}), with the fixed mean frequency
and the variance, but different values of the asymmetry parameter,
in order to see how sensitive the defect induced features of the
reflection spectrum are to the shape of the distribution function.
To this end, we plot the lines of constant height and width for
different values of the asymmetry parameter $n$ ($1 \leq n \leq
2$). The results are presented in Fig.~%
\ref{fig:levels_asymmetry}.

An interesting result revealed by these graphs is that with the
change of the asymmetry, the points at which $W$ and $H$ level
lines cross move parallel to the axis of $\gamma$, while the
respective values of $\Delta$ remain quite stable. This indicates
that the value of $\Delta$, which can be obtained by comparing
experimental reflection spectra with the theory presented in this
paper, is not sensitive to the choice of the distribution function
of the exciton frequencies. The value of the parameter of the
homogeneous broadening is more sensitive to the asymmetry of the
distribution function: it varies by approximately ten percent when
the parameter of the asymmetry changes by a factor of two.
However, the estimate for $\gamma$ can be improved by studying the
temperature dependence of the reflection spectra.

\section{Conclusion}

In the present paper we studied the reflection spectra of a Bragg
multiple quantum well system with a defect: the quantum well at
the center of the structure was replaced by a well with a
different exciton resonance frequency. In an ideal infinite system
such a defect gives rise to a local state with a frequency within
the polariton stop band of the host structure, which reveals
itself in the form of reflection and transmission
resonances.\cite{DefectMQW} The main focus of this paper was on
the effects due to the inhomogeneous broadening, which was taken
into account within the framework of the effective medium
approximation.\cite{KavokinGeneral} Since in
Ref.~\onlinecite{KavokinGeneral} this approach was introduced  on
the basis of qualitative arguments only, in the Appendix we
presented a rigorous derivation of this approximation and
clarified its physical status.

We consider two limiting cases in which defect-induced features of
the spectrum can be described analytically. In one case, the
length of the system is much larger than the penetration length of
the radiation in the infinite periodic structure, $l_c$ (deep
defect). The modification of the reflection in this case is
described by a Lorentz-like minimum, whose depth exponentially
decreases with increasing length of the MQW structure, Eq.~%
(\ref{eq:deep_far}). For long enough structures the presence of
the defect in the structure becomes unnoticeable. A much more
interesting situation arises in the opposite case of systems much
shorter than $l_c$ (shallow defect). In this case, the reflection
spectrum exhibits a sharp minimum followed by a maximum. This
shape of the spectrum resembles a Fano-like resonance observed in
long lossless systems. The position of the maximum of the
reflection is close to the exciton frequency in the defect well,
$\omega_d$, where the short host structure is almost transparent,
and therefore the properties of the spectrum near this point are
similar to those of an isolated quantum well. At the same time,
the minimum is shifted from $\omega_d$ by $\Omega_s = -(\omega_d -
\omega_h)/N$ and due to this shift an influence of the
inhomogeneous broadening, $\Delta$, on the spectrum near this
point is strongly suppressed.  When $\Omega_s \gg \Delta$ the
value of the reflection at the minimum, $R_{min}$, is determined
solely by the small homogeneous broadening, and this results in
extremely small values of $R_{min}$,
Eq.~(\ref{eq:shallow_R_min_max}).

The structure of the spectrum with two extrema persists also in
the case of an intermediate relation between $l_c$ and $N$. This
situation, however, is more complicated and can only be analyzed
numerically. Nevertheless, due to the shift of the minimum
reflection frequency one  can conclude that the minimum value of
the reflection is determined mostly by the homogeneous broadening,
while the distance between the maximum and the minimum is affected
by an additive combination of the homogeneous and inhomogeneous
broadenings. This circumstance allowed us to suggest a simple
method of  extracting both $\gamma$ and $\Delta$ of the defect
well from the reflection spectrum of the structure.

\begin{acknowledgments}
We are grateful to Steven Schwarz for reading and commenting on
the manuscript. The work is supported by AFOSR grant
F49620-02-1-0305 and PSC-CUNY grants.
\end{acknowledgments}

\appendix*
\section{The effective medium approximation for a single quantum well}

The objective of this appendix is to derive ``from  first
principles" the main result of the effective medium approximation,
Eq.~(\ref{eq:app_r0_t0solutions}) with $\kappa$ replaced by the
average susceptibility, given by Eq.~(\ref{eq:inhom_susc}).

A wave incident at a normal, $z$, direction on a quantum well can
be described by a scalar form of Maxwell's equations for one of
the polarizations parallel to the plane of a quantum well:
\begin{equation}\label{eq:app_Maxwell}
  -\nabla^2 E(z) = \frac{\omega^2}{c^2}
    \left(\epsilon_\infty E(z) + 4\pi P(z)\right),
\end{equation}
where $P(z)$ is the polarization due to quantum well excitons. The
latter is determined by an expression similar to Eq.~%
(\ref{eq:polarization}) with the exciton frequency being a random
function of the in-plane coordinate $\rho$
\begin{equation}\label{eq:app_polarization}
  P(z, \omega) = \int d^2\rho\, \chi(\rho, \omega)\Phi(z)\Phi(z')
  E(z', \rho) ,
\end{equation}
where the susceptibility is
\begin{equation}\label{eq:app_susceptibility}
  \chi(\rho, \omega) = \frac{\alpha}{\omega(\rho) - \omega -
  i\gamma}.
\end{equation}
In order to simplify our calculations we make an assumption that
$\Phi(z)$ can be approximated by a delta-function. This
approximation is sufficient for our particular goals here, but the
results obtained will remain valid for more rigorous treatment of
the excitonic wave functions as well. After the Fourier
transformation with respect to the in-plane coordinates,
Eq.~(\ref{eq:app_Maxwell}) can be presented in the form
\begin{eqnarray}\label{eq:app_Fourier_eq}
  -\left(\kappa_q^2 + \frac{d^2}{dz^2}\right) E_x(q,z) \\
  =
  \frac{2\omega^2 \delta(z)}{c^2}\int d^2 q \,\chi(q-q', \omega)
    E(q', z),
\end{eqnarray}
where $\kappa_q^2 = \omega^2\epsilon_\infty/c^2 - q^2$.

Let us represent the susceptibility as a sum of its average value
and a fluctuating part
\begin{equation}\label{eq:app_susc_div}
  \chi(q,\omega) = \langle \chi(\omega) \rangle \delta(q) +
      \tilde \chi(q, \omega),
\end{equation}
where $\langle \chi(\omega) \rangle$ can be written in the form
\begin{equation}\label{eq:app_susc_av}
  \langle \chi(\omega) \rangle = \int d\omega_0
   f(\omega_0)\frac{\alpha}{\omega_0 - \omega - i\gamma}.
\end{equation}
%
The electromagnetic wave existing to the left of the quantum well
consists of the incident and reflected waves. We will see later
that the structure of the Maxwell equations with the polarization,
Eq.~(\ref{eq:app_polarization}), dictates that the reflection
coefficient has a $\delta$-functional singularity in the specular
direction, $q=0$. It is convenient to separate this singularity
from the very beginning and to present the wave at the left-hand
side of the quantum well in the following form.
\begin{equation}\label{eq:app_Em}
 E_-(q,z) = \left(E_0 e^{ikz} + E_0 r_0 e^{-ikz}\right)\delta(q)
 +  E_0 r(q)  e^{-i\kappa_qz}.
\end{equation}
 A similar expression for the wave at the right-hand side
of the quantum well containing only transmitted waves can be
written  as
\begin{equation}\label{eq:app_Ep}
 E_+(\rho,z) = E_0 t_0 e^{ikz} \delta(q)
 +  E_0 t(q) e^{i\kappa_q z}.
\end{equation}

After the substitution of Eqs.~(\ref{eq:app_susc_av}),
(\ref{eq:app_Em}), and (\ref{eq:app_Ep}) into Eq.~%
(\ref{eq:app_Fourier_eq}) we obtain an integral equation for  the
reflection and transmission coefficients, $r(q)$, and $t(q)$.
Assuming that the random process representing the fluctuating part
of $\chi$ does not include constant or almost periodic
realizations, we can conclude that $\tilde\chi(q,\omega)$, does
not have $\delta$-like singularities in almost all realizations.
In this case, the terms proportional to $\delta(q)$ in this
equation must cancel each other independently of other terms. This
leads to a system of equations for $r_0$ and $t_0$ with the
solutions
\begin{equation}\label{eq:app_r0_t0solutions}
  r_0 = \frac{i\eta}{1-i\eta}, \qquad
  t_0 = \frac{1}{1-i\eta},
\end{equation}
where
\begin{equation}\label{eq:app_eff_susc}
  \eta = \frac{\omega}{c\sqrt{\epsilon_\infty}}\langle \chi \rangle.
\end{equation}
These expressions coincide with those of the effective medium
approach, Eq~(\ref{eq:r_and_t}), with an accuracy up to the phase
factor, which does not appear in our derivation because of the
$\delta$-functional approximation for the excitonic wave function.
We can conclude, therefore, that the substitution of the average
susceptibility into the Maxwell equations allows us to conside the
singular contribution to the reflection (transmission) only. The
remaining terms, which are neglected in this approximation, give
only small corrections to the reflection (transmission) in the
specular direction. However, these terms become important when one
considers scattering of light from quantum wells. Another
important assumption in our derivation concerns the spatial
dispersion of the excitons. The existence of the singular
contribution to the reflection (transmission) coefficients is
directly related to neglecting any exciton motion in the in-plane
direction. Spatial dispersion of excitons will smooth out this
singularity, and, therefore, the smoothness of experimental
scattering spectra can be used, in principle, to estimate the
exciton's mass.

It is also important to emphasize that the outlined procedure does
not involve any averaging of the electric field, and therefore the
results obtained describe an intrinsically ``deterministic"
contribution to the reflection and transmission coefficients
rather than some \textit{average} characteristics thereof. The
existence of such a singular component in the scattering spectrum
was mentioned in Ref.~\onlinecite{CitrinScattering} as well as
observed in numerical calculations of
Ref.~\onlinecite{SavonaUltrafast}.

\end{document}